\documentclass[preprint]{elsarticle}

\usepackage{hyperref}
\usepackage{bm}
\usepackage{amsmath}
\usepackage{esdiff}
\usepackage{booktabs}
\usepackage[tight,nice]{units}
\usepackage[T1]{fontenc}
\usepackage[utf8x]{inputenc}
\journal{}

\begin{document}

\begin{frontmatter}
    \title{Collective motion of groups of self-propelled particles following interacting leaders}
\author[biofiz,etho]{B.~Ferdinandy\corref{cor}}
\ead{fbence@elte.hu}
\author[biofiz,statfiz]{K.~Ozogány}
\author[biofiz,statfiz]{T.~Vicsek}

\cortext[cor]{Corresponding author}
\address[biofiz]{Department of Biological Physics, Eötvös Loránd University, H-1117 Pázmány Péter sétány 1/A, Budapest, Hungary}
\address[etho]{MTA-ELTE Comparative Ethology Research Group, Hungarian Academy of Science and Eötvös Loránd University, H-1117 Pázmány Péter sétány 1/C, Budapest, Hungary}
\address[statfiz]{MTA-ELTE Statistical and Biological Physics Research Group, Hungarian Academy of Science and Eötvös Loránd University, H-1117 Pázmány Péter sétány 1/A, Budapest, Hungary}

\begin{abstract}
    In order to keep their cohesiveness during locomotion gregarious animals must make collective decisions. Many species boast complex societies with multiple levels of communities. A common case is when two dominant levels exist, one corresponding to leaders and the other consisting of followers. In this paper we study the collective motion of such two-level assemblies of self-propelled particles. We present a model adapted from one originally proposed to describe the movement of cells resulting in a smoothly varying coherent motion. We shall use the terminology corresponding to large groups of some mammals where leaders and followers form a group called a harem.  We study the emergence (self-organization) of sub-groups within a herd during locomotion by computer simulations. The resulting processes are compared with our prior observations of a Przewalski horse herd (Hortobágy, Hungary) which we use as results from a published case study. We find that the model reproduces key features of a herd composed of harems moving on open ground, including fights for followers between leaders and bachelor groups (group of leaders without followers). One of our findings, however, does not agree with the observations. While in our model the emerging group size distribution is normal,  the group size distribution of the observed herd based on historical data have been found to follow lognormal distribution. We argue that this indicates that the formation (and the size) of the harems must involve a more complex social topology than simple spatial-distance based interactions.

\end{abstract}
\begin{keyword}
    collective motion \sep SPP model \sep collective motion of groups \sep hierarchy

\end{keyword} 
\end{frontmatter}
%\linenumbers
\section{Introduction and motivation}
\label{sec:introduction}

Living in social structures with multiple levels of hierarchy is widespread in the animal kingdom \cite{Grueter2012a,Grueter2012}. Examples range across several taxa, beginning with humans and primates \cite{Abegglen1984,Kummer1968}, through elephants \cite{Wittemyer2005}, to whales \cite{Baird2000,Whitehead2012} and equids \cite{Feh2001,Rubenstein2004}. There are numerous examples of subgroups  forming around a single individual. For example groups may emerge around a matriarch from her descendants, like in african elephants \cite{Wittemyer2005}, sperm whales \cite{Whitehead2012}, and
killer whales \cite{Baird2000}. Alternatively a reproductive unit may form around a breeding male with several breeding females and their young as in Przewalski
horses \cite{Boyd1994} and plains zebras \cite{Rubenstein2004}. These breeding units can sometimes also include non-breeding males as well, like in hamadryas baboons \cite{Kummer1968} or geladas \cite{Dunbar1975}. 

Our aim in the current study was to examine the way in which such a two-level hierarchy may spontaneously emerge in a group and what implications that hierarchy might have for the collective motion of the group. Our motivation and empirical basis was the collective motion of a Przewalski horse herd in light of group formation within the herd, aided by observations made in \cite{ozogany2014} at the Hortobágy National Park in Hungary. As mentioned before, the Przewalski horse herd is split into harems, organized around a breeding male, with several breeding females and their young offspring. So-called bachelor groups, which consist of males that do not have there own harem are also present \cite{hartmann2012}. It should be noted, that although zebra harems form herds in the wild and have a very similar social structure to the Przewalski horse, the Przewalski herd at Hortobágy is only semi-wild as it lives in a bounded environment, which may force them into a herd. Although this has not been studied thoroughly, park officials reported, that the initial population did not form a herd, which only appeared after the growth of population density.

Both the collective motion of several different species of animals \cite{Vicsek2012}, and the emergence of hierarchy within the social system of the Przewalski horse \cite{ozogany2014} have already been modelled. Conversely, the collective motion of animals that are hierarchically organized into subgroups within a larger group have not been modelled. Thus, we aimed to construct a model of group formation and collective motion of a herd composed of sub-groups as a self-propelled particle (SPP) model in two dimensions, where we identified leaders forming harems, and followers making up these harems. 

Leadership is a common concept invoked to explain coordinated group movements. In ungulates this is often attributed to a single individual. A recent study \cite{bourjade2015} raises interesting questions about the validity of such a concept, based on observation of two groups of 12 and 6 Przewalski horses. Using different definitions of leadership (moving first, moving in front, or eliciting joining to movement), no individuals that could be consistently classified as a leader were identified. Some limitations to that study are that several types of movements were not measured. In addition movements in the breeding season were also not measured; this was deemed problematic because in the breeding season the stallions directly elicit movements of their harems away from other stallions. Also, due to methodological reasons, only short periods of the day were observed. It has been shown in some cases that in the same group different type of leadership hierarchies might arise in different contexts \cite{Watts150518,nagy2013context}, and there are examples in nature where certain individuals in animal groups consistently act as leaders, for example in zebras and dolphins \cite{larissa2009,fischhoff2007}. These imply that although attributing leadership to a single individual might not be applicable in all circumstances, it does have explanatory power in a wide range of scenarios. As such it stands to reason that conceptualizing the division between leaders and followers dichotomously helps simplify modelling at a minor cost. Simplifying modelling is helpful in the initial understanding of the type of collective movement we analysed in the present article. Nonetheless this recent study has implications that warrant further field studies of leadership in animals. It would be particularly illuminating to have detailed and continuous data on the movement of large groups of Przewalski horses, which is not easy task, since we do not know of any herd, where attaching measurement devices to the animals is allowed by officials. A substitute for real wild horses could be domestic and feral horses, on which studies have also been carried out \cite{briard2015,krueger2014}. Interestingly, these studies concentrate on movement initiation and not the collective motion itself, showing a somewhat different point-of-view in physicists and biologists.

Herein we consider an earlier SPP model of collective cell movement \cite{szabo2006} and extend it with a two level hierarchy by introducing two distinct types of particles (i.e.\ leaders  and followers) while simultaneously attempting to limit the increase in the number of parameters. In contrast with \cite{ozogany2014} the group formation is not driven by the environment of the herd, but by interactions dynamically evolving during the collective motion of the individuals. While formulating the model, we concentrated on mimicking the movements of Przewalski horses. While this specificity adds some complexity to our model, relative to what is usual in statistical physics, it is mostly related to nuances in movement and does not play a major factor in group formation, which was our main focus. 

Our study could have potential implications for understanding how and why group formation occurs in nature, how group formation affects the system in which it is happening and the rules governing collective motion in a two-level system. Inferring the universalities and the particulars of the different kind of mechanisms, could potentially be used to artificially control both living and human-made systems, such as domestically kept horse herds or flocks of drones.

\section{Model}

\label{sec:model}

\begin{figure}[ht!]

\centering
        \includegraphics[scale=0.5]{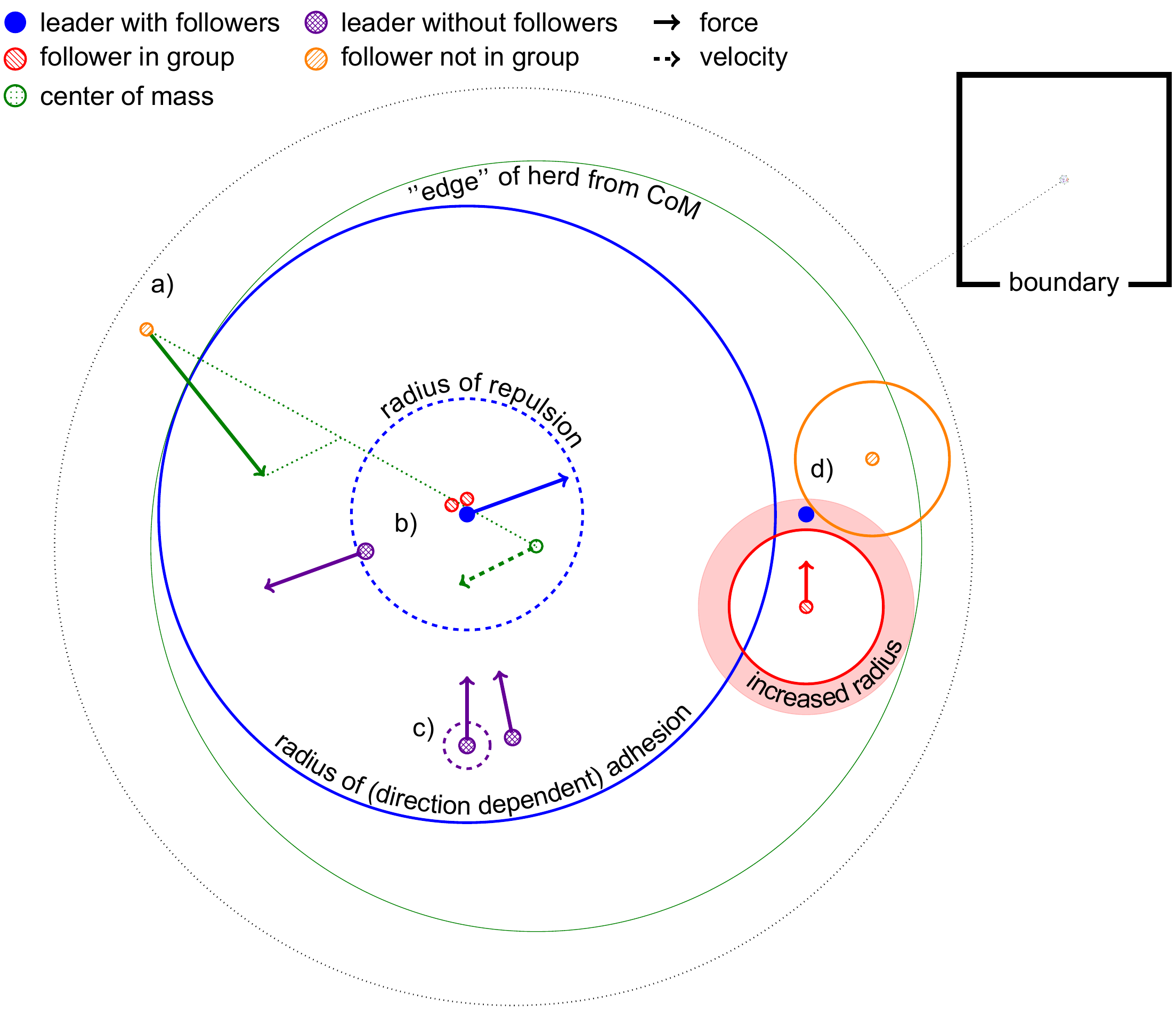}

\caption{Graphical overview of the model depicting a small herd inside the boundary with  various parts of $\bm{F}_\text{int,r}(\bm{r}_i,\bm{r}_j)$ and $\bm{F}_\text{com}$ shown. Radii are drawn to scale (cf. Table \ref{tab:parameters} for actual values), and the herd is magnified from within the boundary to show the forces. Solid arrows depict direction of forces, dashed arrows depict actual velocities. The following details are included: a) a horse farther from the center of mass than the given boundary (large green circle centred on the center of mass) will move towards the herd but also in the direction the herd is going, b) \& c) leaders without groups can go closer to each other than to a leader with a group, while followers can go even closer to a leader, d) the attraction radius of the follower-leader interactions is generally smaller than that of the leader-leader interactions, but it is increased when interacting with the leader of the follower's group.
}%
\label{fig:modeldescription}

\end{figure}

The model is based on \cite{szabo2006} in which a model was developed to depict the collective motion of cells. We modified this model to accommodate two types of SPP-s (leaders and followers), asymmetric interactions and group formation rules. While extending this model we aimed at minimizing the number of extra parameters. Compared with the usual SPP models the model of \cite{szabo2006} gives smoother results due to intrinsic relaxation times.  We choose parameters that allow the development of motion that resembles the movements of a herd made of harems as close as possible within the framework of the model. We provide a graphical overview of the model in Figure \ref{fig:modeldescription} and an introduction here.

The movements of the horses in the model are confined to a square area, large compared to the size of the herd, representing the herding area available to them (Figure \ref{fig:modeldescription} boundary). Periodic boundary conditions were not considered, first, because it is not realistic, and second, because it does not make sense in a co-moving herd to conceptualize that the front may interact with the rear. Also, we introduce a tendency for horses that stray too far from the herd to head back while still going in the general direction of the herd's (Figure \ref{fig:modeldescription} a)).

All horses may follow all other horses, but the strength of the interaction depends on the types and orientations of the SPP-s in question. Given, that it is plausible that leaders must also pay attention to followers, they will follow followers too, but to a much smaller extent than the other way around. Although the interactions taking place are based on metric distances, we introduce a directedness, meaning that a horse will follow the ones in front of it more than the ones behind it. Several types of interaction modes have been suggested in modelling collective motion. Early models used a simple metric distance, e.g., interacting with anybody nearer than a given distance \cite{vicsek1995}. Later topological distances were introduced, e.g., interacting with a fixed number of nearest neighbours \cite{ballerini2008}. Recently it has been proposed that the most biologically correct interaction ranges should be based on visual perception \cite{strandburg2013}. In our case, vision plays little part as equine vision is near 360° \cite{brooks1999} and neither the distances within the herd nor the density of the herd imply that occlusion would have a major effect on interactions. As such, the effect of following the ones in front, rather than the ones behind is related more to the logic of not turning around if there are others heading in the same direction as oneself.

Leaders who acquire followers (i.e., a harem), will stay farther away from other leaders than if they were without followers (Figure \ref{fig:modeldescription} b) and c)). Harems are established based on spatial distance, but followers will gradually belong more and more to the leader they follow, making it easier for them to stay close, because of the stronger and slightly longer distance interactions with their leader than with another leader (Figure \ref{fig:modeldescription} d)).

Our model starts from randomized initial positions and velocities, without followers being assigned to any leader, thus all followers find groups and leaders at the same time. Our model forgoes the introduction of complex social rules by using only spatial interactions as described above and not taking into consideration that in reality, a new horse would be introduced to a herd already split into harems. On the other hand, taking the latter into consideration would not allow for the study of emergent group formation.

\subsection{Formal model description}
\label{sec:formal model}
We have $N_\text{L}$ number of leaders and $N_\text{F}$ number of followers (the list of parameters can be found in Table \ref{tab:parameters}). The 2-dimensional motion of the horse $i \in \lbrace 1$,$ N = N_\text{L} + N_\text{F} \rbrace$ is described by the overdamped dynamics
\begin{equation}
\diff{\bm{r}_i(t)}{t} = v_i^0\bm{n}_i(\theta_i) + 
\sum\limits_{\substack{j=1\\j\neq i}}^N\bm{F}_{\text{int}}(r_{ij},\varphi_{ij}) 
+ \bm{F}_\text{com}(\bar{\bm{r}}-\bm{r}_i,\bar{\bm{v}}) + 
\bm{F}_\text{wall}(\bm{r}_i,\bm{v}_i) + \boldsymbol\xi
\end{equation}

where $t$ is time, $\bm{r}_i$ is the position of and $\bm{v}_i$ is the velocity of horse $i$, $v^0$ is a preferred speed which differs for leaders and followers, $\bm{n}_i$ is a unit vector characterized by the angle $\theta_i$,  $\bm{F}_{\text{int}}$ is a pairwise interaction with $r_{ij} = |\bm{r}_i - \bm{r}_j|$ and $\varphi_{ij}$ being the angle between $\bm{r}_i - \bm{r}_j$ and $\bm{v}_i$, $\bm{F}_\text{com}$ is a global force dependant on the position ($\bar{\bm{r}}$) and the velocity ($\bar{\bm{v}}$) of the center of mass of the herd, $\bm{F}_\text{wall}$ is the force acting at the boundaries and $\boldsymbol\xi$ is a vector whose components are delta-correlated white noise terms with zero mean.

The direction of the self-propelling velocity $\bm{n}_i(t)$, described by the angle $\theta_i(t)$, attempts to relax to $\bm{v}_i(t) = \operatorname{d}\!\bm{r}_i(t)/ \operatorname{d}\!t$ with a relaxation time $\tau_i$:

\begin{equation}
\frac{\operatorname{d}\!\theta_i(t)}{\operatorname{d}\!t} = \frac{1}{\tau_i}\operatorname{arcsin}\!\left[\left( \bm{n}_i(t) \times \frac{\bm{v}_i(t)}{|\bm{v}_i(t)|}\right)\cdot \bm{e}_z\right],
\end{equation}
where $\bm{e}_z$ is a unit vector orthogonal to the plane of motion, and $\tau_i$ differs for leaders and followers. This relaxation provides smooth transitions of the $\bm{n}_i(t)$ desired velocities. The value of $\tau$ was chosen larger for leaders than followers, implying that leaders are harder to ''convince'' than followers to change directions, but our results are not sensitive to changes in $\tau$. 

The $\bm{F}_{\text{int}}(r_{ij},\varphi_{ij})$ force that carries the direct interaction between the horses can be split into the product of a spatial part ($\bm{F}_{\text{int,r}}(r_{ij})$), and a coefficient part ($F_{{\text{int,}\varphi}}(\varphi_{ij}$), the latter being dependent on the angle of the direction of horse $j$ from horse $i$ and the direction of the velocity of horse $i$. The spatial part consists of a pair-wise, asymmetrical force, the direction of which lies on the line passing through the center of masses of the interacting horses and the magnitude of which is the function of the distance $r_{ij}$ between the horses \cite{szabo2006}. The actual form of the force depends on the type of horses involved:

\begin{equation}
\bm{F}_\text{int,r}(r_{ij}) = 
\begin{cases}
\bm{F}_\text{LL}(r_{ij}), & \mbox{if } i \mbox{ and } j \mbox{ are both leaders,}  \\
\bm{F}_\text{FL}(r_{ij}), & \mbox{if } i \mbox{ is a follower and } j \mbox{ is leader,}\\
\bm{F}_\text{LF}(r_{ij}), &\mbox{if } i \mbox{ is a leader and } j \mbox{ is a follower,} \\
\bm{F}_\text{FF}(r_{ij}),& \mbox{if } i \mbox{ and } j \mbox{ are both followers.} 
\end{cases}
\end{equation}

For all four cases there are two radii defined, $R^\text{AT}$ which is the range of attraction, and a smaller radii $R^\text{EX}$, which is the range of repulsion, and also a distance $L$, which defines a distance inside $R^\text{AT}$ but outside of $R^\text{EX}$, splitting the force into four parts depending on distance, namely a repulsive, an attractive and two non-interacting regimes, with different coefficients for all four types of interaction in both the interacting regimes ($F^\text{AT}$ for the attractive and $F^\text{EX}$ for the repulsive), thus having 8 radii with 8 coefficients and 4 distances. On the example of $\bm{F}_\text{LF}(r_{ij})$ the equations look like this (leader-leader and follower-leader interactions are slightly different):

\begin{equation}
\label{eq:sampleforce}
\bm{F}_\text{LF}(\bm{r}_i,\bm{r}_j) = \bm{e}_{ij} \times
\begin{cases}
F^\text{EX}_\text{LF} \frac{r_{ij}-R^\text{EX}_\text{LF}}{R^\text{EX}_\text{LF}}, & r_{ij} < R^\text{EX}_\text{LF},\\
0, & R^\text{EX}_\text{LF} < r_{ij} < R^\text{EX}_\text{LF} + L_\text{LF}, \\

F^\text{AT}_\text{LF} \frac{r_{ij}-R^\text{EX}_\text{LF}}{R^\text{AT}_\text{LF} - R^\text{EX}_\text{LF}-L_\text{LF}}, & R^{EX}_\text{LF} + L_\text{LF}\le r_{ij} \le R^\text{AT}_\text{LF}, \\

0, & R^\text{AT}_\text{LF} < r_{ij},

\end{cases}
\end{equation}
where $\bm{e}_{ij} = (\bm{r}_i -\bm{r}_j)/r_{ij}$. The non-interacting part between $R^\text{EX}$ and $R^\text{AT}$ was chosen to be very small its only function being is to remove some ''vibrations'' that arise at such low densities, when a horse is on the edge of the attractive and repulsive regimes. The form of the force is one of the simplest ways to define gradually growing forces based on distances and the values of the specific parameters were chosen to imitate that leaders with harem wish to protect their followers from other leaders, while bachelor leaders themselves can create groups.

In the cases of leader-leader ($\bm{F}_\text{LL}$) and follower-leader ($\bm{F}_\text{FL}$) interaction this picture is slightly changed due to the formation of groups. Followers will develop a certain amount of affinity to leaders who are close by, that increases in strength when they are close to the leader and decreases when they are farther away from the leader. Each follower keeps track of time spent near each leader with the quantities $D_{ij} \in [0,\infty]$, which follow the simple dynamics
\begin{equation}
\label{eq:maledist}
\frac{\operatorname{d}\!D_{ij}}{\operatorname{d}\!t} = 
\begin{cases}
+1, & r_{ij} \leq R^\text{AT}_\text{LF}, \\
-1, & r_{ij} > R^\text{AT}_\text{LF} \text{ and } D_{ij} > 0, \\
\quad 0, & r_{ij} > R^\text{AT}_\text{LF} \text{ and } D_{ij} \leq 0.
\end{cases}
\end{equation}
This is then translated into an affinity
\begin{equation}
\label{eq:affinity}
A_{ij} = 2A\left(\frac{1}{1+\exp\left(\frac{-D_{ij}}{\tau_A}\right)}-0.5\right)+1,
\end{equation}
where $\tau_A$ is the characteristic time of affinity increase and $A$ is a constant. The form of Eq. \ref{eq:affinity} was chosen so that $A_{ij}$ goes smoothly from $1 \rightarrow A+1$ as $D_{ij}$ goes from $0 \rightarrow\infty $. This effectively changes the parameters in Eq. \ref{eq:sampleforce} (but not in Eq. \ref{eq:maledist}!) for the $\bm{F}_\text{FL}$ case from $F^\text{AT}_\text{FL} \rightarrow A_{ij}F^\text{AT}_\text{FL}$ and from $R^\text{AT}_\text{LF} \rightarrow A_{ij}R^\text{AT}_\text{LF}$. This allows a follower to split farther from the leader it belongs to, without leaving the harem, thus introducing more consistency into the group compositions.

The definition of groups is based on the values $D_{ij}$. Every follower is considered to be in the group of the leader for which the value of $D_{ij}$ is largest for the given follower. The leader--leader interaction differs in one aspect if either of the participating leaders have a group, by effectively increasing the repulsive radius $R^\text{EX}_\text{LL}$ of both leaders fivefold when interacting with each other. As such two leaders can be close to each other only if they don’t each have their own groups. This is reminiscent of the distinction between bachelor groups, where males are close together and harems, where the males are farther apart.

The velocity dependent part is the same for both leaders and followers:
\begin{equation}
F_{\text{int,}\varphi}(\varphi_{ij}) = \frac{-1}{1+\exp(-4(\varphi_{ij}-\frac{\pi}{2}))} + 1,
\end{equation}
which effectively means, that a horse will pay more attention to horses that are in front of it, rather than those that are behind it. The form was chosen because of the saturation properties. The total interaction is thus
\begin{equation}
\bm{F}_{\text{int}}(r_{ij},\varphi_{ij}) = F_{\text{int,}\varphi}(\varphi_{ij})\bm{F}_\text{int,r}(r_{ij}).
\end{equation}

The force $\bm{F}_\text{com}$ keeps the herd roughly together, since if one strays farther than $R_\text{com}$ from the center of mass of the herd it will experience the force 
\begin{equation}
\bm{F}_\text{com}(\bar{\bm{r}}-\bm{r}_i,\bar{\bm{v}}) = 
F_\text{com}\frac{|\bm{r}_i - \bar{\bm{r}}|-R_\text{com}}{R_\text{com}}\left( \frac{\bar{\bm{r}}-\bm{r}_i}{|\bar{\bm{r}}-\bm{r}_i|} +
\beta\frac{\bar{\bm{v}}}{|\bar{\bm{v}}|}
\right),
\end{equation}
where $\beta$ is parameter that tunes how much the horse is guided in the direction the center of mass is heading and $F_\text{com}$ is the overall strength of the force. Since $R_\text{com}$ is relatively large this force is usually inactive, but will smoothly guide a lost horse back into the herd (adopted from \cite{kunal2010}).

The force $\bm{F}_\text{wall}$ sets the boundary conditions. The herd is confined to a square area defined by the length $D$. This box is impenetrable and horses cannot leave it. For the herd to approach this hard boundary in a realistic way, there is a characteristic distance $R_\text{wall}$ where the force $\bm{F}_\text{wall}$ is turned on:
\begin{equation}
\bm{F}_\text{wall}(\bm{r}_i,\bm{v}_i) = \frac{F_\text{wall}}{2}\left( \sin \left[\pi\left( \frac{R_\text{wall}-d_{iw}}{R_\text{wall}}-\frac{1}{2} \right)\right] + 1 \right)
\begin{pmatrix}
\bm{v}_i\cdot\bm{n}_w \\
\bm{v}_i\cdot\bm{t}_w 
\end{pmatrix},
\end{equation}
where $d_{iw}$ is the distance of the horse from the boundary, $\bm{n}_w$ is the normal vector of the boundary and $\bm{t}_w$ is the tangent vector of the boundary, driving the horses smoothly along the wall (adopted from \cite{viragh2014} and \cite{viragh2014corr}).

Initially both leaders and followers are evenly distributed over a square with a linear size of 500, with velocities also randomly distributed.

\subsection{Parameters}
Going, in a na\"{\i}ve way, from the one-type-particle model of \cite{szabo2006} to the two-type-particle model would increase the number of required parameters from 14 to 30 (some parameters are doubled and some are increased fourfold given every possible combination of the particles). By considering that some of these are unnecessary to duplicate (or make four of) our model has 23 parameters. Of these only 7 are relevant in the sense that the formation of meaningful groups is sensitive to their value (parameters that would destroy cohesion even in a one-type-particle model were not taken into account), not considering the size of the herd. A parameter was considered relevant if an increase by twofold or a decrease by half resulted in 0.1\% of followers not being in a group on average (this is less than one per a realization of the model). For a complete list of parameters see Table \ref{tab:parameters}. Parameters were chosen so that cohesive movement occurs and that group formation happens. Except for cases where there was a reason to do otherwise, parameters that could be different for leaders and follower were kept the same. The distances were chosen based on observations, the coefficients of the various forces were chosen so that the phenomenology of the movements resembles that of a real herd. The leaders are slightly faster than followers so that they are able to stay  in front of their harem. It must be noted, that in many cases, leaders in real-life examples may not be at the front of their group, but rather at the side or behind; we elected to use the leading-from-front paradigm for the purpose of simplicity. Other choices pertaining to parameter value selection have been mentioned in the previous section describing the model.

\begin{table}[]
\centering

\begin{tabular}{@{}llll@{}}
\toprule
variable & description & default value & approx. dimensions   \\ \midrule

\multicolumn{4}{c}{relevant variables} \\ \midrule
 $A$ & affinity of followers for leaders & 1.3  & 1.3\\
 $\tau_\text{A}$ & characteristic time of affinity & 500 & \unit[218]{s}\\
  $F_\text{FL}^\text{AT}$ & strength of F-L attraction & 0.03 & \unitfrac[0.0125]{m}{s} \\

   $R_\text{LL}^\text{AT} $ & radius of L-L attraction & 200 & \unit[36]{m} \\
 $R_\text{LL}^\text{EX} $ & radius of L-L repulsion & 15 & \unit[2.7]{m}\\
 $R_\text{LL}^\text{*EX} , 5 R_\text{LL}^\text{EX}$ & -- & 75 & \unit[13.6]{m}\\
 $F_\text{LL}^\text{AT} , F_\text{LF}^\text{AT}$ & strength of L-L and L-F attraction & 0.01 & \unitfrac[0.0042]{m}{s} \\ 

$N_\text{L}$ & number of leaders & 25 & 25 \\
$N_\text{F}$ & number of followers & 175 & 175 \\

\midrule        
\multicolumn{4}{c}{irrelevant variables} \\ \midrule
  $F_\text{FF}^\text{AT}$ & strength of F-F attraction & 0.0002 & \unitfrac[0.000083]{m}{s} \\ 
   $F_\text{LL}^\text{EX}$&  strength of L-L repulsion  & 2 & \unitfrac[0.83]{m}{s} \\ 
$v_\text{L}^{0}$ & velocity of leaders & 1 & \unitfrac[0.416]{m}{s}\\
$v_\text{F}^{0}$ & velocity of followers & 0.9 & \unitfrac[0.375]{m}{s}\\
$\tau_\text{L}$ & L velocity relaxation time & 3 & \unit[1.31]{s}\\
$\tau_\text{F}$ & F velocity relaxation time & 1 & \unit[0.44]{s}\\
 $\xi$& strength of the noise  & 0.5 & \unitfrac[0.21]{m}{s}  \\
 $L_\text{LL} , L_\text{LF} ,L_\text{FL} ,L_\text{FF}$& non-interaction distances & 1 & \unit[0.18]{m}\\
 $R_\text{LF}^\text{AT} , R_\text{FL}^\text{AT}  , R_\text{FF}^\text{AT}$ & radii of attraction & 50 & \unit[9.1]{m}\\
 $R_\text{LF}^\text{EX}  , R_\text{FL}^\text{EX} , R_\text{FF}^\text{EX}$ & radii of repulsion & 5 & \unit[0.9]{m}\\
$F_\text{LF}^\text{EX}  , F_\text{FL}^\text{EX} , F_\text{FF}^\text{EX}$&  strength of repulsion   & 5 & \unitfrac[2.08]{m}{s} \\  
  $R_\text{com}$ & radius of the cohesion force & 250 & \unit[45.5]{m}\\
 $F_\text{com}$ & strength of the cohesion force & 2.5 &\unitfrac[1]{m}{s} \\
 $\beta$ & cohesion force parameter & 0.01 & 0.01\\
 $F_\text{wall}$ & strength of boundary repulsion & 3 & \unitfrac[1.25]{m}{s}\\
 $R_\text{wall}$ & distance of boundary repulsion & 200 & \unit[36.4]{m}\\
 $D$ & linear size of bounding box & 10000 & \unit[1800]{m}\\
 \bottomrule
\end{tabular}

\caption{Table of the parameters of the model grouped according to relevancy in group formation. L and F abbreviate leader and follower respectively. The approximate proper dimensions are based on a comparison with observed horses, see Section \ref{sec:dimscales} for details.}\label{tab:parameters}
\end{table}

\section{Results}

\label{sec:results}

Our model, with the given parameters, produces a cohesive and ordered motion of the entire herd, while forming groups around leaders and also bachelor groups from group-less leaders. This is in qualitative agreement with the actual observed herd moving on an open plane and as an interesting extra phenomenon, our model also includes ``fights'' between leaders for followers. By ``fights'' we mean a situation where two or more leaders without groups get extremely close to one or more followers and after a short time, one of the leaders ``wins'', i.e. a follower is ascribed to be in the leader's group for long enough for it to chase away the other leaders (see Supplementary video 1).  

We find that the forming of groups within the herd causes cohesiveness to drop compared to a case without groups. We also find, that in accordance with but with a greater precision than the previous study, the group size distribution of the horses living in the Hortobágy National Park is lognormal. In contrast to this, the current model, based solely on spatial interactions, gives a normal distribution, which implies that spatial interactions alone are not enough to produce the observed group structure.

\subsection{Cohesiveness of the herd}
\begin{figure}
\centering
\includegraphics[scale=1]{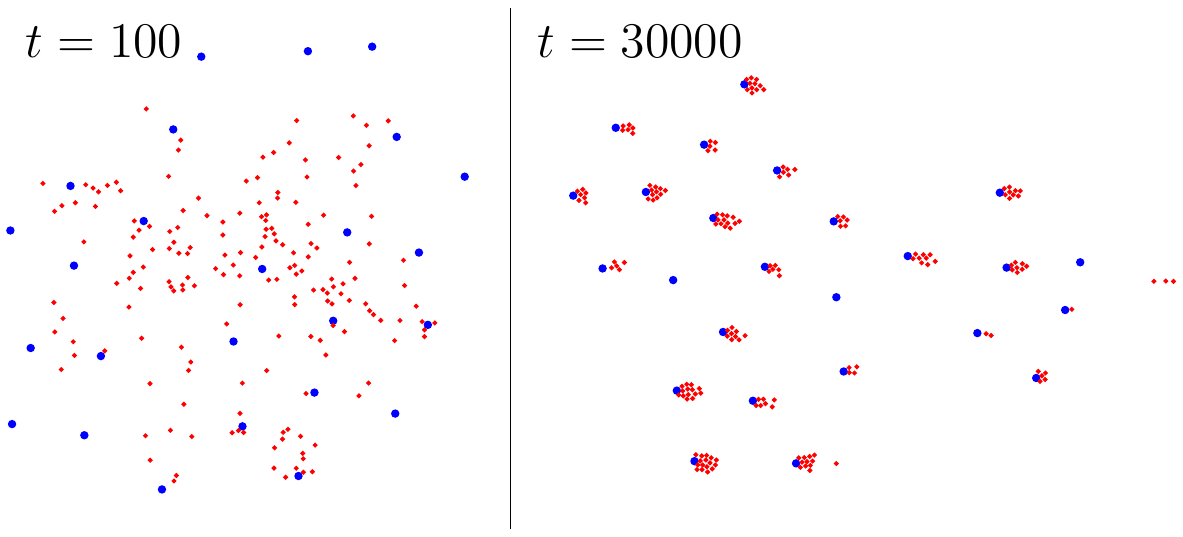}%
\caption{Starting from a uniform random distribution of positions and velocities (left side) the herd forms groups and exhibits ordered motion (right side). Blue dots represent leaders and red dots represent followers (see Supplementary video 2 for a video example).}
\label{fig:screenshot}
\end{figure}

Starting from uniform random initial positions and velocities of the individuals, after sufficient time, the model develops ordered motion throughout the herd while forming groups and thus arriving at a structured and co-moving herd (see Figure \ref{fig:screenshot} and Supplementary video 2). We assume that during collective migration the horses cannot stop, thus there are two phases of ordered movement:  translational movement, and collective rotation about the -- otherwise slowly moving -- center of mass. Indeed, when it is not possible to stop (e.g. due to fear), but is not feasible or desirable to move the herd as a whole, herding mammals have been observed to rotate around a common point. To measure translational cohesiveness we use the following translational order parameter

\begin{equation}
\Phi_t = \frac{1}{N} \left |\sum_{i = 1}^{N} \frac{\bm{v}_i}{|\bm{v}_i|} \right |,
\end{equation}

and to measure the rotational cohesiveness we introduce the following rotational order parameter

\begin{equation}
\Phi_c = \frac{1}{N}\sum_{i = 1}^N \mathcal{P}\left(\frac{\bm{v}_i}{|\bm{v}_i|}\right),
\end{equation}

where $\mathcal{P}$ denotes projection onto the normal of the line going through $\bm{r}_i$ and $\bar{\bm{r}}$.

Going from a totally disordered translational movement to totally ordered translational movement $\Phi_t$ will grow from 0 to 1, while $\Phi_c$ will move from -1 to 1, as the system moves from a totally ordered rotation around the center of mass in one direction, through no collective rotation to totally ordered rotation in the other direction.

\begin{figure}
\centering
\includegraphics[scale=0.65]{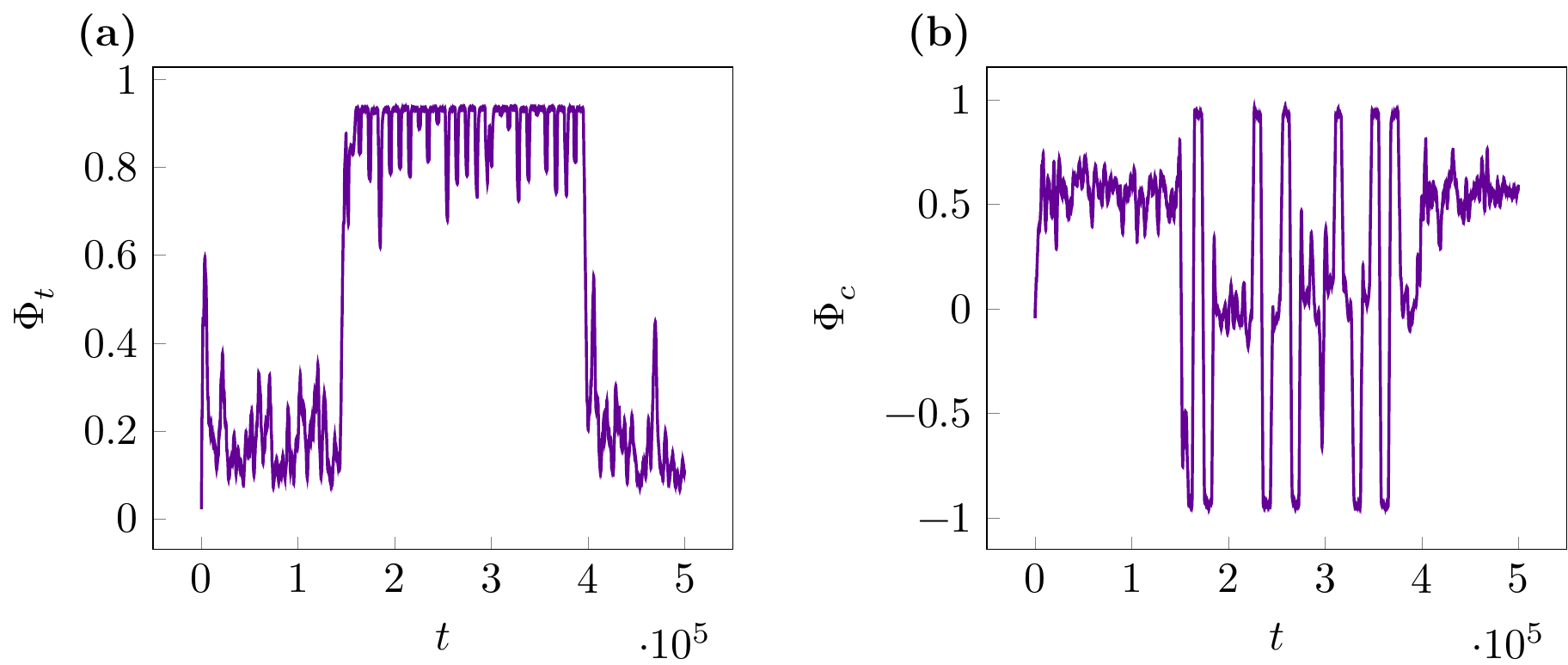}%
\caption{The herd as a whole either exhibits an ordered translational motion or rotates around a slowly drifting center of mass. This two different type of motion can be distinguished due to the values of the translational (a) and rotational (b) order parameters. The plots are from the same specific run of the model, with the curves smoothed by a window of $\Delta t = 1000$. The spikes during the translational phase are caused by the confining wall (see Supplementary video 4 for a sample of an interaction with the wall).}
\label{fig:motiontypes}
\end{figure}

We find that the system, with parameters given in Table \ref{tab:parameters} switches between two modes, one of ordered rotation and one of ordered translational motion (see Figure \ref{fig:motiontypes} and Supplementary video 3 for an example of a transition from rotational to translational motion). Since the horses in the model do not have the capacity to stop, in an event of indecision about the direction to move they must rotate about a common axis, namely the center of mass. By averaging over the full length of 1000 runs in total, we find that the rotations have no specific direction, as expected (Mann-Whitney U-test, $p=0.96$ on left-right similarity).

\begin{figure}[ht]
\centering
        \includegraphics[scale=1]{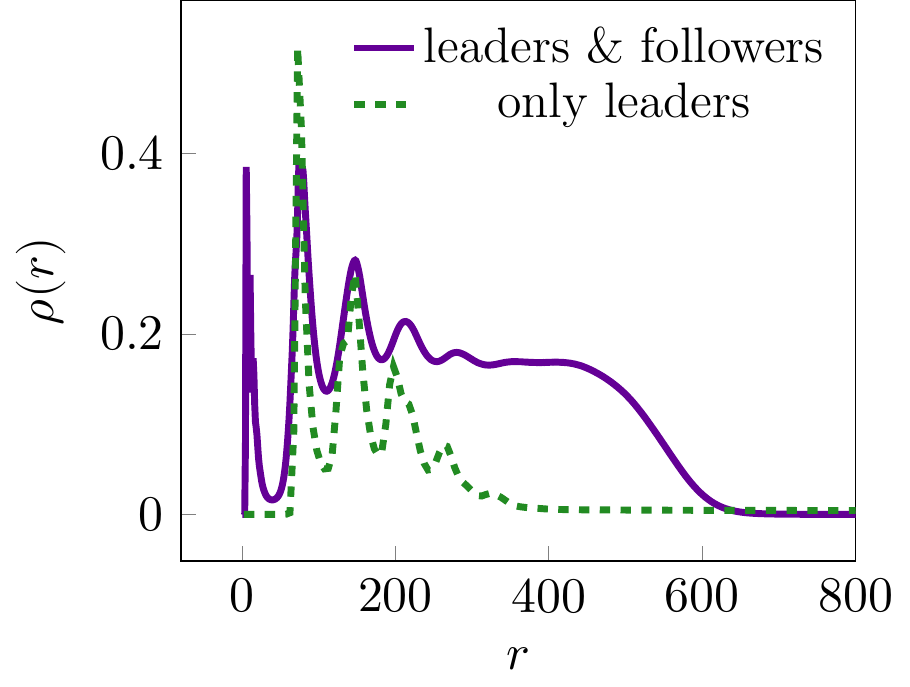}

\caption{%
The pair-correlation for a herd composed of leaders and followers, but only calculated on the leaders (solid line), and in the case where only leaders are present (striped line). In the latter case the distances are scaled with $R_{LL}^{*EX}/R_{LL}^{EX}$, to compensate for the effect of no leader having a group (leaders without followers can be closer to each other than ones with followers). The main structure of the herd, even with followers, is set by the leaders, but the presence of followers slackens the rigidity of this structure.
}
\label{fig:densitycorr}
\end{figure}

Calculating the pair-correlation function

\begin{equation}
\rho(\bm{r}) = \langle \delta(\bm{r}-\bm{r}_i) \rangle,
\end{equation}
for the leaders in the normal scenario (i.e. where followers are present) and in the scenario where followers are missing, we find that the main structure of the herd is given by the leaders, and introducing followers only slightly loosens this (aside from the fact that it increases the distances between the leaders, see Figure \ref{fig:densitycorr}). We also investigated the effect of introducing followers among the leaders on the order parameter of the translational movement. Comparing $\Phi_t$ (calculated using only the velocities of the leaders in two cases, one where there are only leaders and one where there are also followers) we find that order is decreased when allowing for followers and forming of groups (see Table \ref{tab:orderparam}). This loss in the efficiency of the movement of the herd as a whole points to benefits gained from social groupings outside the paradigm of simple locomotion.

\begin{table}[ht!]
\centering

\begin{tabular}{@{}lllll@{}}

\toprule
&  $\langle\Phi_t\rangle$ & $\langle\Phi_t\rangle$ (only leaders) & $\langle\Phi_c\rangle$ &$\langle \Phi_c\rangle$ (only leaders) \\
\midrule
%short leaders & 0.745 & 0.745 & 0.010 & 0.010  \\
%short all & 0.567 & 0.609 & 0.000 & 0.000 \\
w/o followers & $0.866 \pm 0.016$ & -- & $-0.002 \pm 0.011$ & -- \\
with followers & $0.608 \pm 0.018$ & $0.633 \pm 0.017$& $0.025 \pm 0.020$ & $0.026 \pm 0.021$  \\
%shortmanymales & 0.802 & 0.802 & 0 & 0  \\
\bottomrule

\end{tabular}

\caption{The translational and rotational order parameters averaged over 120 simulations with standard errors. The duration of the runs were many times longer than the stabilization of groups. The first row is from simulations where only leaders were present, the second row is the full model with followers. In this case the averages were calculated on the whole herd as well as on the leaders only. Adding followers and thus moving in groups decreases the order of translational movement, implying that group formation has  benefits other than increased herd cohesion. Although the herd would rotate often, as  expected, there is no specific direction of the rotation (Mann-Whitney U-test on 1000 runs, where the simulations was terminated at a time not long after stabilization of groups yields a $p = 0.96$ on left-rigth similarity).}\label{tab:orderparam}

\end{table}

\subsection{Group size distribution}
\begin{figure}[ht]
\centering
        \includegraphics[scale=1]{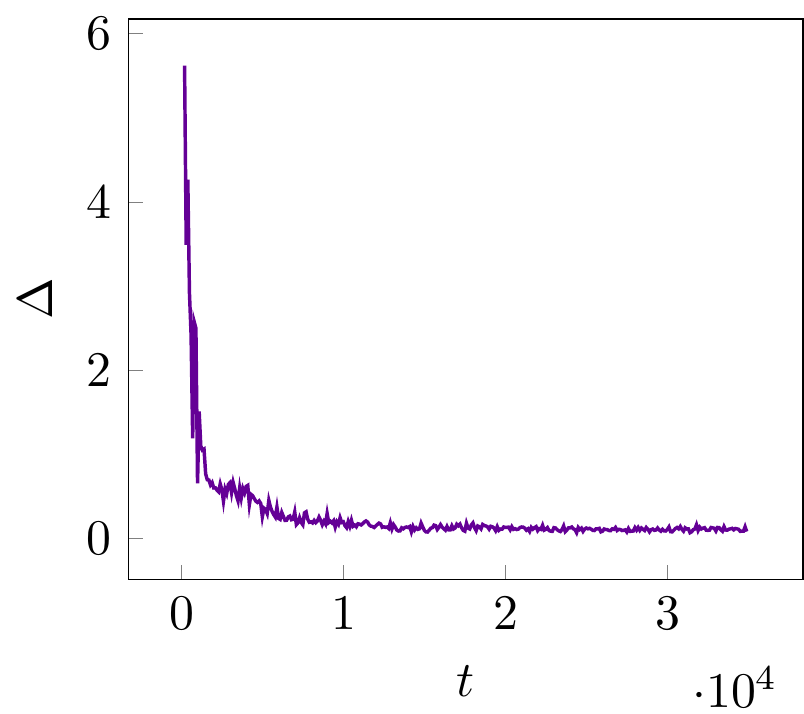}

\caption{%
The groups size distribution quickly stabilizes as it is shown by the plot of $\Delta$. To calculate $\Delta$ we create a histogram of the group sizes at regular intervals during a simulation and take the sum of the differences of each respective bin of the histogram in two consecutive measurements. Each point is averaged over 1000 independent simulations. 
}
\label{fig:haremdistributiontime}
\end{figure}

Starting from a uniform random spatial distribution and group-less state, the model, after 
sufficient time, will produce co-moving groups based on the relative positions of leaders and followers. The emerging group size distribution is normal, although some leaders and followers may not belong to a group. We define groups by the highest (non-zero) $D_{ij}$ values of the followers, i.e a group consists of the leader and the followers with their highest $D_{ij}$ rating corresponding to this leader. This effectively means that groups are formed by followers spending the most time with a specific leader.  The group size distribution rapidly reaches a close-to-final state and after some time relaxes to the final state (see Figure \ref{fig:haremdistributiontime}). We show the transition by creating a histogram of the group sizes at regular intervals during a simulation and taking the sum of the differences of each respective bin of the histogram in two consecutive measurements, averaged over 1000 independent simulations .

\begin{figure}[ht]
\centering
        \includegraphics[scale=1]{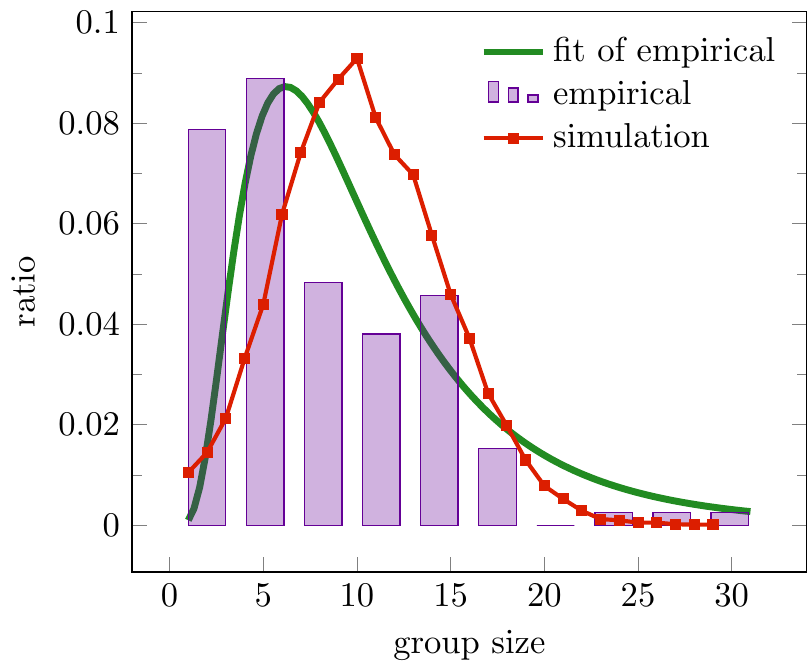}

\caption{Comparison of the group size distribution in the model with an empirical one (the group size distribution of Przewalski horses living in the Hortobágy National Park). The empirical distribution follows a clearer lognormal distribution than in \cite{ozogany2014} due to the incorporation of historical data. The distribution obtained from the model is close to normal and is mean-fitted to the empirical distribution. We attribute the difference to the fact that the social interactions of horses are too complex to capture in purely spatial interactions.
}
\label{fig:haremdistribution}

\end{figure}

On Figure \ref{fig:haremdistribution} we show a comparison of the simulated distribution with real data obtained from a Przewalski horse herd (see \cite{ozogany2014} for details). Since harem sizes gradually change over time among the horses, the real data has been improved by taking into account historical harem size distributions, showing a more clear lognormal distribution than in the previous study of \cite{ozogany2014}. In this previous study a network model was formulated to account for the lognormal distribution of the group sizes, while the current model, based on purely spatial interactions was not able to reproduce this. This indicates that at this level of complexity, it is not possible to reduce social interactions to spatial interactions.

\subsection{Dimension scales}
\label{sec:dimscales}
Horses usually travel by walking, which is roughly around $\unitfrac[1.5]{km}{h}$ based on our aerial observations averaged over several minutes. In this model $v^0_L$ is the corresponding parameter of the walking speed. To compare our model's length scale with that of reality we have calculated the pair-correlation function of the of the wild horses by using aerial pictures of the real herd and that of the herd in our model and compared the first peaks. This roughly equates the arbitrary length unit of our model to $\unit[0.18]{m}$ in reality (c.f. $R_\text{LF}^{EX}$ in Table  \ref{tab:parameters}). From this we can calculate that the arbitrary time unit of our model is roughly equal to $\unit[0.44]{s}$%0.43636363636363634
. This puts $\tau_\text{L}$ and $\tau_\text{F}$ at about reaction time ($\unit[0.5-1.5]{s}$), $\tau_\text{A}$ to about 3 and a half minutes, and the emergence of a coherent collective motion, with stable harems to slightly less than $\unit[10]{minutes}$. Since $\tau_\text{L}$ and $\tau_\text{F}$ both characterize a fast cognitive process it is not unrealistic that the characteristic times are on the scale of reaction times. Since in wild horses the groups do not form from randomly distributed individuals spontaneously, but rather evolve in an already laid down social context, the time needed for group formation is not readily comparable to that of the real herd. On the other hand, for a group of 200 unfamiliar individuals, where leaders are already appointed and everybody is actually already moving, the 10 minutes seems like a reasonable time for group formation (the authors' personal experience with spontaneous group formation in human groups of comparable sizes would allow for even longer times).

\section{Conclusions}

\label{sec:discussion}
As the only truly wild horse in the world, the Przewalski horses, now mostly living in relatively easily accessible nature reserves, have drawn considerable attention.  Both their collective movements \cite{bourjade2009} and the formation of their harems have attracted interest \cite{ozogany2014}.
However, the unique collective motion displayed by this species, as a large herd consisting of cohesive harems moving together in a coordinated way, has not been modelled to date.

Our model, adapted from a model designed for cells, is  able to qualitatively reproduce the motion of a wild horse herd moving on an open plain, along with formation of groups consisting of one leader and some followers and bachelor groups (group of leaders without followers), with a roughly adequate correspondence of dimension scales. During the analysis of the behaviour of the model we found three interesting phenomena.

First, the herd in our model will at times rotate around its center of mass, while we have not observed the horses to circle, many animals do. This is the direct effect of the fact, that in our model the individuals are unable to stop. Indeed, animals that do rotate along a common axis are usually also unable to stop (e.g.\ flying animals) or is infeasible or dangerous for them to stop. Although some efforts have already been made to model the stopping of a group of animals \cite{kunal2010,Ferdinandy2012}, we suggest further investigations into a model, that would allow for not only the stopping of, but also for the resuming of locomotion.

Second, the translational order parameter is decreased when we introduce followers among the leaders, thus the considered grouping process within the herd effectively reduces locomotion efficiency. In many systems the interactions during motion that give rise to collective motion is for the sake of more efficient locomotion of the group as a whole, but the harem formation within a herd is first and foremost due to reproductive reasons. Thus it is not surprising that the reproductive benefits might outweigh the slight decrease in locomotive efficiency. 

Third, the results obtained from our model are not in agreement with the observed group size distribution of the herd that motivated our work (the latter being a lognormal while the former being a normal distribution). Our simple model operates solely with interactions based on spatial distances, while group-forming processes in real societies have many complex attributes, thus deviations from the exact features of the empirical population is expected. On the other hand, the collective motion in many species can be described by purely distance-based interactions, making the exact nature of these deviations non-trivial. Consequently, we propose further investigations of collectively moving systems to find the properties that allow for the spatial formulation of interactions within the system. It can be supposed that in systems where individuals are interchangeable (in the meaning that individual recognition during the motion is not feasible), like a group of cells, ants or a flock of starlings, considering only distance-based interactions is enough to reproduce the observed collective motion pattern, but in animals living in structured social systems (and maintaining an individual recognition), like horses, social factors are much more important during interactions than actual distances, thus interchangeability might be one such property.

\section{Acknowledgements}
We are grateful to the Directorate of Hortobágy National Park for providing us with their data as well as authorizing our observations. This work was supported by the Hungarian Academy of Sciences (grant numbers MTA 01 031 and 2011TKI552).

\section*{References}
\bibliography{refs}

\begin{thebibliography}{10}
\expandafter\ifx\csname url\endcsname\relax
  \def\url#1{\texttt{#1}}\fi
\expandafter\ifx\csname urlprefix\endcsname\relax\def\urlprefix{URL }\fi
\expandafter\ifx\csname href\endcsname\relax
  \def\href#1#2{#2} \def\path#1{#1}\fi

\bibitem{Grueter2012a}
C.~C. Grueter, B.~Chapais, D.~Zinner,
  \href{http://dx.doi.org/10.1007/s10764-012-9618-z}{Evolution of multilevel
  social systems in nonhuman primates and humans.}, Int J Primatol 33~(5)
  (2012) 1002--1037.
\newblock \href {http://dx.doi.org/10.1007/s10764-012-9618-z}
  {\path{doi:10.1007/s10764-012-9618-z}}.
\newline\urlprefix\url{http://dx.doi.org/10.1007/s10764-012-9618-z}

\bibitem{Grueter2012}
C.~C. Grueter, I.~Matsuda, P.~Zhang, D.~Zinner,
  \href{http://dx.doi.org/10.1007/s10764-012-9614-3}{Multilevel societies in
  primates and other mammals: Introduction to the special issue.}, Int J
  Primatol 33~(5) (2012) 993--1001.
\newblock \href {http://dx.doi.org/10.1007/s10764-012-9614-3}
  {\path{doi:10.1007/s10764-012-9614-3}}.
\newline\urlprefix\url{http://dx.doi.org/10.1007/s10764-012-9614-3}

\bibitem{Abegglen1984}
J.~Abegglen, On Socialization in Hamadryas Baboons: a field study., Bucknell
  University Press, Lewisburg, 1984.

\bibitem{Kummer1968}
H.~Kummer, Social Organisation of Hamdryas Baboons. A Field Study., University
  of Chicago Press, Chicago, 1968.

\bibitem{Wittemyer2005}
G.~Wittemyer, I.~Douglas-Hamilton, W.~Getz,
  \href{http://www.sciencedirect.com/science/article/pii/S0003347205000667}{The
  socioecology of elephants: analysis of the processes creating multitiered
  social structures}, Animal Behaviour 69~(6) (2005) 1357 -- 1371.
\newblock \href
  {http://dx.doi.org/http://dx.doi.org/10.1016/j.anbehav.2004.08.018}
  {\path{doi:http://dx.doi.org/10.1016/j.anbehav.2004.08.018}}.
\newline\urlprefix\url{http://www.sciencedirect.com/science/article/pii/S0003347205000667}

\bibitem{Baird2000}
R.~Baird, The killer whale: Foraging specializations and group hunting., in:
  J.~Mann, R.~Connor, P.~Tyack, H.~Whitehead (Eds.), Cetacean Societies,
  University of Chicago Press, Chicago, 2000, p. 127–153.

\bibitem{Whitehead2012}
H.~Whitehead, R.~Antunes, S.~Gero, S.~Wong, D.~Engelhaupt, L.~Rendell,
  \href{http://dx.doi.org/10.1007/s10764-012-9598-z}{Multilevel societies of
  female sperm whales (physeter macrocephalus) in the atlantic and pacific: Why
  are they so different?}, International Journal of Primatology 33~(5) (2012)
  1142--1164.
\newblock \href {http://dx.doi.org/10.1007/s10764-012-9598-z}
  {\path{doi:10.1007/s10764-012-9598-z}}.
\newline\urlprefix\url{http://dx.doi.org/10.1007/s10764-012-9598-z}

\bibitem{Feh2001}
C.~Feh, B.~Munkhtuya, S.~Enkhbold, T.~Sukhbaatar,
  \href{http://www.sciencedirect.com/science/article/pii/S0006320701000519}{Ecology
  and social structure of the gobi khulan equus hemionus subsp. in the gobi b
  national park, mongolia}, Biological Conservation 101~(1) (2001) 51 -- 61.
\newblock \href
  {http://dx.doi.org/http://dx.doi.org/10.1016/S0006-3207(01)00051-9}
  {\path{doi:http://dx.doi.org/10.1016/S0006-3207(01)00051-9}}.
\newline\urlprefix\url{http://www.sciencedirect.com/science/article/pii/S0006320701000519}

\bibitem{Rubenstein2004}
D.~Rubenstein, M.~Hack, Natural and sexual selection and the evolution of
  multi-level societies: insights from zebras with comparisons to primates.,
  in: P.~Kappeler, C.~van Schaik (Eds.), Sexual Selection in Primates: New and
  Comparative Perspectives, Cambridge University Press, New York, 2004, p.
  266–279.

\bibitem{Boyd1994}
L.~Boyd, K.~Houpt, Przewalski's Horses, State University of New York Press, New
  York, 1994.

\bibitem{Dunbar1975}
R.~Dunbar, P.~Dunbar, Social dynamics of gelada baboons., Contrib Primatol 6
  (1975) 1--157.

\bibitem{ozogany2014}
K.~Ozogány, T.~Vicsek,
  \href{http://dx.doi.org/10.1007/s10955-014-1131-7}{Modeling the emergence of
  modular leadership hierarchy during the collective motion of herds made of
  harems}, Journal of Statistical Physics (2014) 1--19\href
  {http://dx.doi.org/10.1007/s10955-014-1131-7}
  {\path{doi:10.1007/s10955-014-1131-7}}.
\newline\urlprefix\url{http://dx.doi.org/10.1007/s10955-014-1131-7}

\bibitem{hartmann2012}
E.~Hartmann, E.~Søndergaard, L.~J. Keeling,
  \href{http://www.sciencedirect.com/science/article/pii/S0168159111003091}{Keeping
  horses in groups: A review}, Applied Animal Behaviour Science 136~(2–4)
  (2012) 77 -- 87.
\newblock \href
  {http://dx.doi.org/http://dx.doi.org/10.1016/j.applanim.2011.10.004}
  {\path{doi:http://dx.doi.org/10.1016/j.applanim.2011.10.004}}.
\newline\urlprefix\url{http://www.sciencedirect.com/science/article/pii/S0168159111003091}

\bibitem{Vicsek2012}
T.~Vicsek, A.~Zafeiris,
  \href{http://www.sciencedirect.com/science/article/pii/S0370157312000968}{Collective
  motion}, Physics Reports 517~(3–4) (2012) 71 -- 140, collective motion.
\newblock \href
  {http://dx.doi.org/http://dx.doi.org/10.1016/j.physrep.2012.03.004}
  {\path{doi:http://dx.doi.org/10.1016/j.physrep.2012.03.004}}.
\newline\urlprefix\url{http://www.sciencedirect.com/science/article/pii/S0370157312000968}

\bibitem{bourjade2015}
M.~Bourjade, B.~Thierry, M.~Hausberger, O.~Petit,
  \href{http://dx.doi.org/10.1371/journal.pone.0126344}{Is leadership a
  reliable concept in animals? an empirical study in the horse}, PLoS ONE
  10~(5) (2015) 1--14.
\newblock \href {http://dx.doi.org/10.1371/journal.pone.0126344}
  {\path{doi:10.1371/journal.pone.0126344}}.
\newline\urlprefix\url{http://dx.doi.org/10.1371/journal.pone.0126344}

\bibitem{Watts150518}
I.~Watts, B.~Pettit, M.~Nagy, T.~B. de~Perera, D.~Biro,
  \href{http://rsos.royalsocietypublishing.org/content/3/1/150518}{Lack of
  experience-based stratification in homing pigeon leadership hierarchies},
  Royal Society Open Science 3~(1).
\newblock \href
  {http://arxiv.org/abs/http://rsos.royalsocietypublishing.org/content/3/1/150518.full.pdf}
  {\path{arXiv:http://rsos.royalsocietypublishing.org/content/3/1/150518.full.pdf}},
  \href {http://dx.doi.org/10.1098/rsos.150518}
  {\path{doi:10.1098/rsos.150518}}.
\newline\urlprefix\url{http://rsos.royalsocietypublishing.org/content/3/1/150518}

\bibitem{nagy2013context}
M.~Nagy, G.~V{\'a}s{\'a}rhelyi, B.~Pettit, I.~Roberts-Mariani, T.~Vicsek,
  D.~Biro, Context-dependent hierarchies in pigeons, Proceedings of the
  National Academy of Sciences 110~(32) (2013) 13049--13054.

\bibitem{larissa2009}
L.~C. David~Lusseau, \href{http://www.jstor.org/stable/40295396}{The emergence
  of unshared consensus decisions in bottlenose dolphins}, Behavioral Ecology
  and Sociobiology 63~(7) (2009) 1067--1077.
\newline\urlprefix\url{http://www.jstor.org/stable/40295396}

\bibitem{fischhoff2007}
I.~R. Fischhoff, S.~R. Sundaresan, J.~Cordingley, H.~M. Larkin, M.-J. Sellier,
  D.~I. Rubenstein, Social relationships and reproductive state influence
  leadership roles in movements of plains zebra, equus burchellii, Animal
  Behaviour 73~(5) (2007) 825--831.

\bibitem{briard2015}
L.~Briard, C.~Dorn, O.~Petit,
  \href{http://dx.doi.org/10.1111/eth.12402}{Personality and affinities play a
  key role in the organisation of collective movements in a group of domestic
  horses}, Ethology 121~(9) (2015) 888--902.
\newblock \href {http://dx.doi.org/10.1111/eth.12402}
  {\path{doi:10.1111/eth.12402}}.
\newline\urlprefix\url{http://dx.doi.org/10.1111/eth.12402}

\bibitem{krueger2014}
K.~Krueger, B.~Flauger, K.~Farmer, C.~Hemelrijk,
  \href{http://www.sciencedirect.com/science/article/pii/S0376635713002222}{Movement
  initiation in groups of feral horses}, Behavioural Processes 103 (2014) 91 --
  101.
\newblock \href
  {http://dx.doi.org/http://dx.doi.org/10.1016/j.beproc.2013.10.007}
  {\path{doi:http://dx.doi.org/10.1016/j.beproc.2013.10.007}}.
\newline\urlprefix\url{http://www.sciencedirect.com/science/article/pii/S0376635713002222}

\bibitem{szabo2006}
B.~Szab\'o, G.~J. Sz\"oll\"osi, B.~G\"onci, Z.~Jur\'anyi, D.~Selmeczi,
  T.~Vicsek, \href{http://link.aps.org/doi/10.1103/PhysRevE.74.061908}{Phase
  transition in the collective migration of tissue cells: Experiment and
  model}, Phys. Rev. E 74 (2006) 061908.
\newblock \href {http://dx.doi.org/10.1103/PhysRevE.74.061908}
  {\path{doi:10.1103/PhysRevE.74.061908}}.
\newline\urlprefix\url{http://link.aps.org/doi/10.1103/PhysRevE.74.061908}

\bibitem{vicsek1995}
T.~Vicsek, A.~Czir{\'o}k, E.~Ben-Jacob, I.~Cohen, O.~Shochet, Novel type of
  phase transition in a system of self-driven particles, Physical review
  letters 75~(6) (1995) 1226.

\bibitem{ballerini2008}
M.~Ballerini, N.~Cabibbo, R.~Candelier, A.~Cavagna, E.~Cisbani, I.~Giardina,
  V.~Lecomte, A.~Orlandi, G.~Parisi, A.~Procaccini, et~al., Interaction ruling
  animal collective behavior depends on topological rather than metric
  distance: Evidence from a field study, Proceedings of the national academy of
  sciences 105~(4) (2008) 1232--1237.

\bibitem{strandburg2013}
A.~Strandburg-Peshkin, C.~R. Twomey, N.~W. Bode, A.~B. Kao, Y.~Katz, C.~C.
  Ioannou, S.~B. Rosenthal, C.~J. Torney, H.~S. Wu, S.~A. Levin, et~al., Visual
  sensory networks and effective information transfer in animal groups, Current
  Biology 23~(17) (2013) R709--R711.

\bibitem{brooks1999}
D.~E. Brooks, A.~Matthews, Equine ophthalmology, Veterinary ophthalmology 2
  (1999) 1165--1274.

\bibitem{kunal2010}
K.~Bhattacharya, T.~Vicsek, Collective decision making in cohesive flocks, NEW
  JOURNAL OF PHYSICS 12.
\newblock \href {http://dx.doi.org/10.1088/1367-2630/12/9/093019}
  {\path{doi:10.1088/1367-2630/12/9/093019}}.

\bibitem{viragh2014}
C.~Virágh, G.~Vásárhelyi, N.~Tarcai, T.~Szörényi, G.~Somorjai, T.~Nepusz,
  T.~Vicsek, \href{http://stacks.iop.org/1748-3190/9/i=2/a=025012}{Flocking
  algorithm for autonomous flying robots}, Bioinspiration \& Biomimetics 9~(2)
  (2014) 025012.
\newline\urlprefix\url{http://stacks.iop.org/1748-3190/9/i=2/a=025012}

\bibitem{viragh2014corr}
C.~Virágh, G.~Vásárhelyi, N.~Tarcai, T.~Szörényi, G.~Somorjai, T.~Nepusz,
  T.~Vicsek, \href{http://stacks.iop.org/1748-3190/9/i=4/a=049501}{Corrigendum:
  Flocking algorithm for autonomous flying robots (2014 bioinspir. biomim. 9
  [http://dx.doi.org/10.1088/1748-3182/9/2/025012] 025012 )}, Bioinspiration \&
  Biomimetics 9~(4) (2014) 049501.
\newline\urlprefix\url{http://stacks.iop.org/1748-3190/9/i=4/a=049501}

\bibitem{bourjade2009}
M.~Bourjade, B.~Thierry, M.~Maumy, O.~Petit,
  \href{http://dx.doi.org/10.1111/j.1439-0310.2009.01614.x}{Decision-making in
  przewalski horses (equus ferus przewalskii) is driven by the ecological
  contexts of collective movements}, Ethology 115~(4) (2009) 321--330.
\newblock \href {http://dx.doi.org/10.1111/j.1439-0310.2009.01614.x}
  {\path{doi:10.1111/j.1439-0310.2009.01614.x}}.
\newline\urlprefix\url{http://dx.doi.org/10.1111/j.1439-0310.2009.01614.x}

\bibitem{Ferdinandy2012}
B.~Ferdinandy, K.~Bhattacharya, D.~Ábel, T.~Vicsek, Landing together: How
  flocks arrive at a coherent action in time and space in the presence of
  perturbations, Physica A 391~(4) (2012) 1207--1215.
\newblock \href {http://dx.doi.org/doi:10.1016/j.physa.2011.10.010}
  {\path{doi:doi:10.1016/j.physa.2011.10.010}}.

\end{thebibliography}

\end{document}